\documentclass[aps,prb,showpacs,amsmath,twocolumn,amssymb,superscriptaddress,letterpaper]{revtex4}
\usepackage{mathrsfs}
\usepackage{graphicx}
\usepackage{graphics}
\usepackage{subfigure}
\usepackage{bm}
\usepackage{dcolumn}
\usepackage{amsmath,bm}

\newcommand{\nd}{{\vphantom{\dagger}}}

\usepackage{color}
\begin{document}
\title{Positive Magneto-conductivity of  Weyl Semimetals in the Ultra-quantum Limit }

\author{Chui-Zhen Chen}
\affiliation{International Center for Quantum Materials, School of Physics,
Peking University, Beijing 100871, China}
\affiliation{Collaborative Innovation Center of Quantum Matter, Beijing, 100871, China}
\author{Haiwen Liu}
\affiliation{International Center for Quantum Materials, School of Physics,
Peking University, Beijing 100871, China}
\affiliation{Collaborative Innovation Center of Quantum Matter, Beijing, 100871, China}
\author{Hua Jiang}
\affiliation{College of Physics, Optoelectronics and Energy,
Soochow University, Suzhou 215006, China}
\author{X. C. Xie}
\affiliation{International Center for Quantum Materials, School of Physics,
Peking University, Beijing 100871, China}
\affiliation{Collaborative Innovation Center of Quantum Matter, Beijing, 100871, China}
\date{\today}

\begin{abstract}
In this paper, we numerically study the magnetic transport properties of  disordered Weyl semimetals (WSM) in the ultra-quantum limit.
We find a positive magnetic conductivity for the long-range disorder, although  the system tends to have negative magnetic conductivity
for the weak short-range disorder.
Remarkably, for long-range disorder, such a positive magnetic conductivity cannot be described by  the semiclassical Boltzmann transport theory even in the weak disorder limit, and the back-scattering assisted by the high Landau levels is always important.
Our results have significant implications for the positive magnetic conductivity recently discovered in the WSM systems,
and point out two physical mechanisms: (i) the long-range correlated disorder suppresses the back-scattering among the zeroth Landau level modes; (ii) with increasing the magnetic field, the back-scattering assisted by the high Landau levels will also be suppressed.
\end{abstract}


\maketitle

\section{ Introduction}
Topological semimetals are novel
states of quantum matter with the gapless bulk states and the nontrival surface states.
Weyl semimetals (WSMs)  are the most celebrated ones which
have attracted considerable attention  recently.\cite{Balents2011,Wan2011,Yang2011,Burkov2011,Xu2011,Bulmash2014,Weng2015,Xu2015,lv2015}
They have Weyl nodes (linear touching points of conduction and valence bands) in the bulk, and Fermi arcs on the surface.\cite{Balents2011,Wan2011,Yang2011}
The WSM phase has been theoretically predicted in the pyrochlore iridates  A$_2$Ir$_2$O$_7$, the magnetic topological insulator multilayers,
as well as noncentrosymmetric transition-metal monophosphides in condensed matter physics.\cite{Wan2011,Yang2011,Burkov2011,Xu2011,Bulmash2014,Weng2015}
Now, the WSM has been realized in TaAs family materials.\cite{Xu2015,lv2015}

The chiral anomaly is one of the most attractive magnetic transport features in the WSMs.\cite{Alder1969,Nielsen1981}
When the magnetic field and electric field are both parallel to a pair of Weyl nodes,
the charge will pump from one Weyl node to the other resulting in non-conservation of chiral charges.\cite{Nielsen1981}
As a consequence, the magnetic transport properties provide the vital information for the chiral anomaly.
Currently, there is a number of theoretic and experimental researches on the magnetic transport behaviors in the WSM and Dirac-semimetal materials.\cite{Hosur2012,Son2013,Abanin2014,Burkov2014,Gorbar2014,Lu2015,LZ2015,Goswami2015,Jia2015,Huang2015,Xiu2015,Ong2015,Kim2013,Xiong2015,Yu2015}
The chiral anomaly is  related to the $B^2$ positive magnetic conductivity of the WSMs for a weak magnetic field $B$,\cite{Son2013,Burkov2014,Jia2015} which may have been observed experimentally.\cite{Huang2015,Xiu2015,Ong2015,Kim2013,Xiong2015,Yu2015}
On the other hand, there is strong dispute about relation between the chiral anomaly
and the magnetic conductivity for a strong magnetic field.\cite{Gorbar2014,Lu2015,LZ2015,Goswami2015}
However, most of the current theoretical studies are based on the semiclassical transport theory or the perturbation theory.
The exact numerical investigations of the magnetic transport properties of the WSMs are desirable.

In this paper, we numerically study the magnetic transport properties of the disordered WSMs under a strong magnetic field $B$ along the Weyl nodes. We investigate both cases with the short-range and long-range disorders, because of their strike differences
in inter-node scattering and Landau levels broadening behaviors.\cite{Xie1988,Xie1990}
For the weak short-range disorder, the system exhibits a negative  magnetic conductivity for the zeroth Landau level, consistent to the semiclassical Boltzmann transport theory. On the other hand, for the long-range disorder, we discover a positive magnetic conductivity.
Remarkably, because the long-range disorder can suppress the inter-node scattering and give rise to a finite Landau level broadening,
the high Landau levels (HLLs) assisted transport are dominant even under a weak disorder and the Boltzmann transport theory for the zeroth Landau level is not applicable.
 We find that  the positive magnetic conductivity origins from two physical mechanisms.
(i) In the presence of the long-range correlated disorder, the scattering among the highly degenerated  zeroth Landau level modes are suppressed in the ultra-high magnetic field.
(ii) The HLLs-assisted backscattering decreases with an increasing magnetic field.

The rest of the paper is organized as follows. In Sec.II,
we introduce a disordered WSM Hamiltonian under the magnetic field and then give
the details of our numerical methods.
In Sec. III, we provide a semiclassical Boltzmann transport theory of the magnetic WSMs
and give the magnetic conductivity of the zeroth Landau level.
In Sec. IV, we show the numerical results of  magnetic transport properties of the WSMs.
In Sec. V, a brief discussion and summary are present.

\section{model Hamiltonian and methods}
\begin{figure}
\centering
\includegraphics[scale=0.55, bb = 0 170 650 480, clip=true]{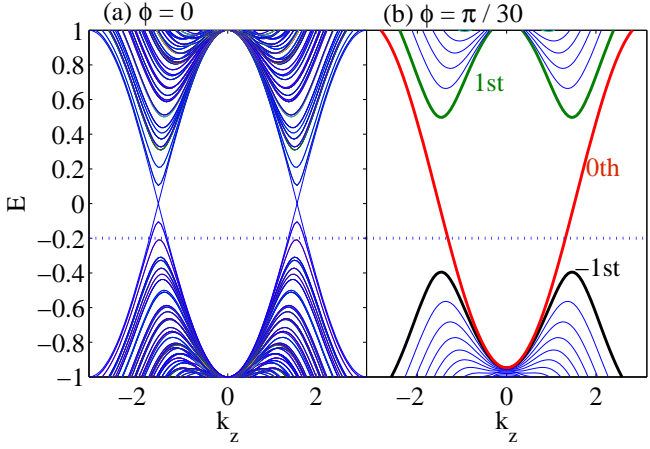}
\caption{(Color online). The dispersion of the Weyl-semimetal Hamiltonian at the different phases (a) $\phi=0$,
and (b) $\phi=\pi/30$, respectively. In (b), the $\pm1$st Landau levels and $0$th Landau level are labeled by bold lines.
The model parameters are $t_z=m_0=v=1$ and $m_z=0$, with the Weyl nodes at $k^{0}_z=\pm\pi/2$.
The Fermi energy $E_F=-0.2$ is labeled by the blue dashed line.
The phase $\phi$ is in the unit of $\phi_0=\hbar/e$.
\label{fig1} }
\end{figure}
We consider a two-band model WSM  Hamiltonian written  as \cite{Yang2011,Chen2015,Lu2015}
\begin{eqnarray}
  H_{0} &=& \Delta_{z}\sigma_{z}  + vk_{x} \sigma_{x}+vk_{y} \sigma_{y},
  \label{equ:1}
\end{eqnarray}
with $\Delta_z=(t_{z}-m_{z})-t_zk_z^2/2  - m_{0}(k_{x}^2+ k_{y}^2)/2$.
Here  $v$, $t_z$, $m_{z}$, $m_{0}$ are model parameters,  and  $\sigma_{x,y,z}$ are Pauli matrices.
The energy spectrum is
\begin{eqnarray*}
  E_{k}&=&\pm\sqrt{\Delta_z^2+(v k_x)^2 +(v  k_y)^2}
\end{eqnarray*}
and there are two (gapless) Weyl nodes at $(k^0_x,k^0_y,\pm k^0_z)=(0,0,\pm\sqrt{2-2m_z/t_z})$, if $m_z/t_z<1$.
In the presence of a magnetic field along $z$ direction, the magnetic Hamiltonian can be obtained by the Pierls substitution
\begin{eqnarray}
  \mathscr{H}_0 &=& M_{z}\sigma_{z}- \frac{m_{0}}{2}[(k_{x}-eBy/\hbar)^2+ k_{y}^2]\sigma_z \nonumber\\
   &&+ v (k_{x}-eBy/\hbar) \sigma_{x}+v k_{y} \sigma_{y}
  \label{equ:2}
\end{eqnarray}
with a magnetic field $B$.
The zeroth Landau level energy is
\begin{eqnarray}
  E_{0k_z}&=&(m_z-t_z)+\frac{t_z}{2}k_z^2+\frac{m_0}{2l_B^2}
  \label{equ:3}
\end{eqnarray}
 and
$\pm n$th Landau level energies are
\begin{eqnarray}
  E_{\pm nk_z}&=& \frac{m_0}{2l_B^2}\pm\sqrt{\frac{2nv^2}{l_B^2}+(\frac{nm_0}{l_B^2} -M_z)^2}
\end{eqnarray}
with $M_z=(t_z-m_z)-t_zk_z^2/2$ and the magnetic length $l_B=\sqrt{\hbar/eB}$.

Now, the clean magnetic  Hamiltonian  $\mathscr{H}_0$ can be discretized into a cubic lattice, \cite{Jiang2012}
resulting
\begin{eqnarray}
  H_{B} &=& \sum_{n} -\psi_n^{\dagger}(m_{z}+2m_0)\sigma_{z}\psi_n^\nd
  + \{\psi_n^{\dagger}\frac{t_z}{2}\sigma_{z}\psi_{n+\hat{z}}^\nd\nonumber\\
  && +\psi_n^{\dagger}(\frac{m_0}{2}\sigma_{z}-\frac{iv}{2}\sigma_{x})\exp(i\phi_{n,n+\hat{x}})\psi_{n+\hat{x}}^\nd\nonumber\\
  &&+\psi_n^{\dagger}(\frac{m_0}{2}\sigma_{z}-\frac{iv}{2}\sigma_{y})\psi_{n+\hat{y}}^\nd\}+h.c.
\end{eqnarray}
where the phase $\phi_{n,m}=\int_{n}^{m}{\bf A}\cdot d{\bf l}/\phi_0$ with the vector potential
${\bf A}=(-By,0,0)$ and $\phi_0=\hbar/e$. $\hat{x}$ ($\hat{y}$, $\hat{z}$) is unit vector in $x$ ($y$, $z$) direction.
$\psi_n^{\dagger}=(\varphi_{n,\uparrow}^{\dagger},\varphi_{n\downarrow}^{\dagger})$ creates a spin-($\uparrow$,$\downarrow$) electron at the $n$th site. The lattice constant is set as $a=1$.
Then we add an anisotropic long-range disorder potential $V({\bf r})$
to the discretized Hamiltonian $ H_{B}$,  and the disordered magnetic Hamiltonian becomes
\begin{eqnarray}
 H&=& H_{B} +  V({\bf r})I_{2\times2}
\end{eqnarray}
with
\begin{eqnarray*}
V({\bf r})&=&\sum_{i=1}^{n} V_i\exp[-\frac{({\bf r_{\|}}-{\bf r}_{i\|})^2}{2\xi_{\|}^2}-\frac{({\bf r}_{z}-{\bf r}_{iz})^2}{2\xi_z^2}]
\end{eqnarray*}
$\sum_{i=1}^{n}$ sums over $n$ impurities randomly located among $N$ lattices at $\{{\bf r}_1,{\bf r}_2,...{\bf r}_n\}$ with ${\bf r}_i=({\bf r}_{\|},{\bf r}_z)$ and $V_{i}$ is uniformly distributed in $[-W/2$,$W/2]$ with $W$ the disorder strength.\cite{Rycerz2007,Zhang2009}
We define the impurity density $n_i=n/N$ for convenience.
The impurity potential ranges in $x$-$y$ plane and $z$ direction are $\xi_\|$ and $\xi_z$, respectively.
When $\xi_\|=\xi_z$, $V({\bf r})$ recovers the isotropic longed-range disorder potential
and it is reduced to short-range (Anderson) disorder if $\xi_\|=\xi_z\rightarrow0$.

In our numerical calculations, we first study the two-terminal conductance of
the sample (in $z$ direction) with the periodical boundary condition in the $x$ and $y$ directions. The system consists of a disordered central region sized $L_x\times L_y \times L_z$ coupling to two clean semi-infinite leads.
The intrinsic conductance is given by $G=[G^{-1}_{LB}-N^{-1}]^{-1}$ with $N$ the number of modes in the leads and the Landauer-B\"{u}ttiker conductivity $G_{LB}$.\cite{Braun1997,Slevin2001,Zhang2009}
$G_{LB}$ is evaluated by Landauer-B\"{u}ttiker formula $G_{LB}=e^2/h {\rm Tr}[\Gamma_L G^{r}\Gamma_R G^{a}]$,\cite{Data1995}
where $G^{r}(E_F)=[G^{a}(E_F)]^{\dagger}=[E_F- H^{cen} -\sum_{n=L,R}\Sigma_n]^{-1}$ is the retarded/advanced Green's function and $\Gamma_{L/R}=i(\Sigma^r_{L/R}-\Sigma^a_{L/R})$
with $H^{cen}$ the Hamiltonian of the central region and $\Sigma_{n=L,R}$ the self-energy of the left/right lead.
For convenience, we define the intrinsic conductivity as $\sigma=G\frac{L_z}{L_x L_y}$.
We also investigate the  density of states $N(E)=\frac{1}{N}\sum_{n=1}^{N}\delta(E-E_n)$ correspondingly  by the kernel polynomial method,\cite{beta2006}  where $E_n$ is the $nth$ eigenvalue of  the $N\times N$ disordered Hamiltonian matrix.
The model parameters are $t_z=m_0=v=1$ and $m_z=0$.
The magnetic field $B$ is expressed in terms of the magnetic phase $\phi=Ba^2/\phi_0$ with $\phi_0=\hbar/e$ and the lattice constant $a=1$.
The Fermi energy are fixed at $E_F=-0.2$ in our numerical simulations.

\section{semiclassical Boltzmann transport theory}

Before turn to numerical simulations, we first investigate magneto-conductivity of the zeroth Landau level by the semiclassical Boltzmann transport theory in the ultra-quantum limit.
In presence of a strong magnetic field along $z$-direction, the system has highly degenerated modes along $z$-direction
for each Landau level [see Fig.~\ref{fig1}(b)].
Under the relaxation time approximation of the linearized Boltzmann equation,\cite{Mahan}
the conductivity can be evaluated by
\begin{eqnarray}
\sigma(E_F)&=&e^2 N_{0}(E_F)v_F^2\tau(k_F),
\label{equ:5}
\end{eqnarray}
and agrees with Einstein relation with the one-dimensional diffusion constant $D=v_F^2\tau(k_F)$.
Here $\tau(k_F)$ is the transport lifetime at Fermi vector $k_F$,
$v_F=\sqrt{2t_z[E_F+t_z-m_z-m_0/(2l_B^2)]}/\hbar$ is the Fermi velocity along $z$ direction,
and $N_{0}(E_F)=1/(2\pi^2\hbar v_Fl_B^2)$ is the density of states of the clean system at the Fermi surface.
It is worth noting that, for fixed Fermi energy $E_F$, the Fermi velocity $v_F$ decreases with the magnetic field $B$,
while the density of states $N_{0}(E_F)$ increases with $B$.
Now the transport lifetime is give by
\begin{eqnarray*}
 \frac{1}{\tau}&=&\frac{2\pi}{\hbar}\sum_{k'_x,k'_z}W_{k_x,k_z;k'_x,k'_z}\delta(E_{0 k_z}-E_{0 k'_z})(1-\cos\theta_{k_z,k'_z})
\end{eqnarray*}
where $\cos\theta_{k_z,k'_z}=\frac{k_zk'_z}{|k_z||k'_{z}|}$ and $W_{k_x,k_z;k'_x,k'_z}=\langle|\langle k_z, k_x, 0|V({\bf r})|0, k_x', k_z' \rangle|^2\rangle$ with $|0, k_x, k_z \rangle$ eigenvector of the zeroth Landau level, $V({\bf r})$ the disorder potential
and $\langle ...\rangle$ averaging over the different disorder configurations.
In the following, we will consider the conductivity and the transport lifetime for different types of disorder.

(i) For the short-range disorder $\langle V({\bf r}_1)V({\bf r}_2)\rangle=\frac{W^2}{12}\delta({\bf r}_1-{\bf r}_2)$,
the inverse transport lifetime and the conductivity are given by
\begin{eqnarray}
  \label{equ:8}\frac{1}{\tau} &=&\frac{W^2}{12}\frac{1}{\pi \hbar^2 v_F(B) l_B^2}\\
  \label{equ:9} \sigma&=& \frac{e^2}{h}\frac{12\hbar^2}{W^2}v_F^2(B)
\end{eqnarray}
where the magnetic length $l_B=\sqrt{\hbar/eB}$  and $v_F$ is the Fermi velocity.
$W$ is the disorder strength  and  $\langle ...\rangle$ averages over different disorder configurations. Since $v_F$ decreases with $B$, the magnetic conductivity $\frac{d\sigma}{dB}$ is negative\cite{Lu2015}.

(ii-a) For the long-range disorder $\langle V({\bf r}_1)V({\bf r}_2)\rangle=\frac{W^2}{12a^3}n_i\exp\{-\frac{({\bf r_{1\|}}-{\bf r}_{2\|})^2}{2\xi_{\|}^2}-\frac{({\bf r}_{1z}-{\bf r}_{2z})^2}{2\xi_z^2} \}$, $\tau$ and $\sigma$ are given by
\begin{eqnarray}
  \label{equ:10}
  \frac{1}{\tau} &=& \frac{W^2\eta_{eff}}{12}\frac{\exp(-2k_F^2\xi_z^2)}{\pi \hbar^2v_F}\frac{1}{\xi_{||}^2+l_B^2}\\
  \label{equ:11}
  \sigma&=& \frac{e^2}{h}\frac{12\hbar^2 v_F^2}{W^2\eta_{eff}}\exp(2k_F^2\xi_z^2)\frac{\xi_{\|}^2+l_B^2}{l_B^2}
\end{eqnarray}
where $\xi_{||}$ and $\xi_z$ are potential correlated length in $x$-$y$ plane and $z$-direction, respectively. $n_i$ is the impurities density
and $\eta_{eff}=n_i2\pi\sqrt{2\pi}\xi_{||}^2\xi_{z}/a^3$ with the lattice constant  $a=1$.

(ii-b) For the anisotropic disorder $\langle V({\bf r}_1)V({\bf r}_2)\rangle\!=\!\frac{W^2n_i}{12a^2}\exp[-\frac{({\bf r_{1\|}}-{\bf r}_{2\|})^2}{2\xi_{\|}^2}] \delta({\bf r}_{1z}-{\bf r}_{2z})$, $\tau$ and $\sigma$ are
\begin{eqnarray}
  \label{equ:12}\frac{1}{\tau} &=&\frac{W^2\eta_{eff}}{12} \frac{1}{\pi \hbar^2 v_F} \frac{1}{\xi_{||}^2+l_B^2},\\
  \label{equ:13}\sigma&=& \frac{e^2}{h}\frac{12\hbar^2 v_F^2}{W^2\eta_{eff}}\frac{\xi_{\|}^2+l_B^2}{l_B^2}.
\end{eqnarray}
where $\eta_{eff}=n_i 2\pi\xi_{||}^2/a^2$ and  the lattice constant  $a=1$.

(ii-c) For the anisotropic disorder case  $\langle V({\bf r}_1)V({\bf r}_2)\rangle\!=\!\frac{W^2n_i}{12a}\exp[-\frac{({\bf r}_{1z}-{\bf r}_{2z})^2}{2\xi_{z}^2}]\delta({\bf r_{1\|}}-{\bf r}_{2\|})$, $\tau$ and $\sigma$ can be expressed as
\begin{eqnarray}
  \label{equ:14}\frac{1}{\tau} &=& \frac{W^2\eta_{eff}}{12}\frac{\exp(-2k_F^2\xi_z^2)}{\pi \hbar^2 v_F}\frac{1}{l_B^2},\\
  \label{equ:15}\sigma&=& \frac{e^2}{h}\frac{12\hbar^2 v_F^2}{W^2\eta_{eff}}\exp(2k_F^2\xi_z^2).
\end{eqnarray}
with $\eta_{eff}=n_i\sqrt{2\pi}\xi_{z}/a$.

Given the same $\eta_{eff}$, the (ii-a) turns into  (ii-b) and (ii-c) if $\xi_z\rightarrow0$ and $\xi_\|\rightarrow0$, respectively,
while (ii-a,b,c) all become (i) if both $\xi_z\rightarrow0$ and $\xi_\|\rightarrow0$.

When the disorder is smooth enough in the $x$-$y$ plane, i.e. $ \xi_{\|}\gg l_B$,
we find the conductivity $\sigma$ is determined by the density of states $N_0(E_F)=(1/2\pi^2\hbar v_Fl_B^2)$, increasing with $B$ [see Eq.(\ref{equ:13})].
Because the degenerated modes in the zeroth Landau level are nearly independent, $\tau$ becomes approximately independent of $B$ [see Eq.~(\ref{equ:12})].
For the smooth disorder along $z$ direction ($\xi_z \gg 1/k_F$),
Eq.~(\ref{equ:15}) indicates that the conductivity should monotonously decrease with increasing $B$ since $v_F$ decreases with $B$.
However, it is crucial that the large $k_z$ backscattering is suppressed by the damping factor $\exp(-2k_F^2\xi_z^2)$, resulting a long transport lifetime [see Eq.~(\ref{equ:14})]. Thus the conductivity should  be unrealistic large (see Eq.\ref{equ:15}).
In this circumstance, the HLLs-assisted backscattering   could become dominant
and thus the result by Eq.(\ref{equ:15}) is no longer applicable.
As a consequence, although  magnetic conductivity is positive in Eq.~(\ref{equ:11}), the result is not quantitatively accurate
and the  HLLs-assisted backscattering is significant.

In the following, we will study the magneto-conductivity of the system for all four types of the disorder. We choose the impurity density
$n_i$ such that the system are influenced by the same effective disorder fluctuation.

\section{Numerical results}

Now let's come to investigate the magnetic transport properties along $z$ direction  numerically.
\subsection{Positive Magneto-conductivity}
\begin{figure}
\centering
\includegraphics[scale=0.32, bb = 0 15 1000 600, clip=true]{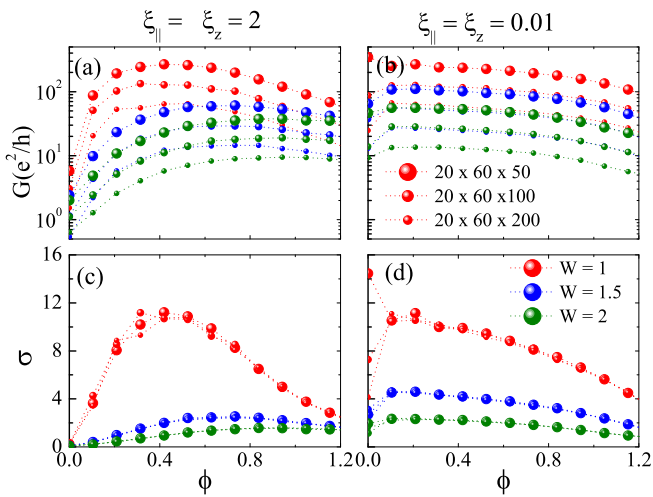}
\caption{(Color online). The conductance $G$ versus the magnetic phase $\phi$  at different disorder strength $W$ with (a) long-range disorder $\xi_{\|}=\xi_{z}=2$ and $n_i=0.02$, and (b) short-range disorder $\xi_{\|}=\xi_z=0.01$ and $n_i=1$, respectively.
(c)-(d) plot conductivity $\sigma=GL_z/(L_xL_y)$ versus $\phi$ with same data as (a) and (b), respectively.
For a given $W$, the conductivity of the different sized samples ($L_x\times L_y\times L_z = 20\times 60 \times 50$, $ 20\times 60 \times 100$, and $20\times 60 \times 200$) collapses to  one line. The Fermi energy $E_F=-0.2$ and the phase $\phi$ is in the unit of $\phi_0=\hbar/e$.
\label{fig2} }
\end{figure}
\begin{figure}
\centering
\includegraphics[scale=0.32, bb =00 0 1000 590, clip=true]{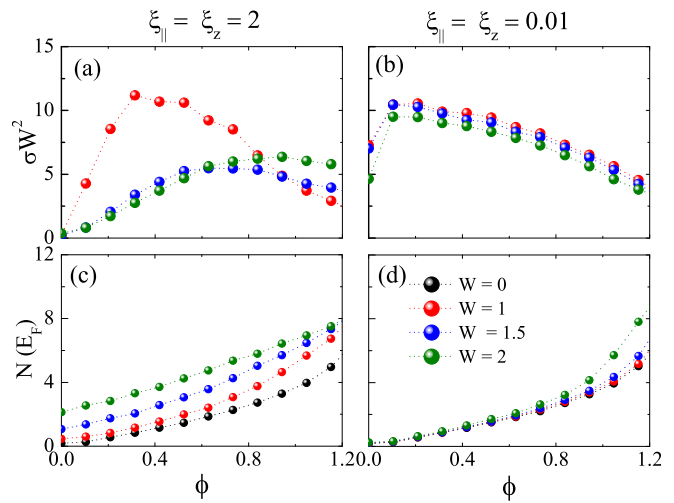}
\caption{(Color online).
(a)-(b) plot $\sigma W^2$ against the magnetic phase $\phi$ for (a) the long-range disorder $\xi_{\|}=\xi_{z}=2$ and (b) the short-range disorder $\xi_{\|}=\xi_{z}=0.01$, respectively. The data are the same as those of Fig.~\ref{fig2}(a) and (b) with size $20\times60\times100$.
(c)-(d) the density of states $N(E_F)$ versus the magnetic phase $\phi$ at different disorder strength, corresponding to two types of
disorder of (a) and (b), respectively. The Fermi energy is set as $E_F=-0.2$ and $\phi$ is in the unit of $\phi_0=\hbar/e$.
\label{fig3} }
\end{figure}
In Fig.~\ref{fig2} (a) and (b), we show the conductance $G$ versus the magnetic phase $\phi$ for the long-range disorder ($\xi_{\|}=\xi_{z}=2$, $n_i=0.02$) and short-range disorder ($\xi_{\|}=\xi_{z}=0.01$, $n_i=1$), respectively.
We find that the conductivity $\sigma$ of all samples for the different disorder strengths collapse to one line respectively, justifying the diffusive transport of the system (for $\phi\neq0$) [see Fig.~\ref{fig2} (c) and (d)].
In Fig.~\ref{fig2} (c), the conductivity $\sigma$ first increases and then decreases with the magnetic phase $\phi$ for long-range disorder.
According to the above Boltzmann transport theory, it seems like that the peak of the conductivity $\sigma$ for the long-range disorder comes from the interplay between the increasing density of states $N_0(E_F)$ and the decreasing the Fermi velocity $v_F$ with $\phi$.
On the contrary, in Fig.~\ref{fig2} (d), the conductivity $\sigma$  decreases monotonously  with $\phi$ for the short-range disorder,
which comes from the decreasing the Fermi velocity $v_F$ with $\phi$ as we discussed in (i) of Sec.~III.

In order to verify the semiclassical Boltzmann transport theory,  we plot $\sigma W^2$ against magnetic phase $\phi$ in Fig.\ref{fig3} (a) and (b). Because $\sigma W^2 (\varpropto \sigma/\tau =e^2N_0(E_F)v_F^2)$ is independent of the disorder strength $W$ [see Eq.~(\ref{equ:9}),(\ref{equ:11}), (\ref{equ:13}) and (\ref{equ:15})], all lines for different $W$ should merge.
In the Fig.~\ref{fig3}(b), all lines with different disorder strengths are very close,
indicating that the Boltzmann approximation applies.
However, in the Fig.~\ref{fig3}(a), the peak of the  conductivity moves to higher magnetic field with increasing disorder strength.
Thus the Boltzmann approximation is strongly violated for the long-range disorder case.
Moreover, we find that the density of states is strongly deviated from the clean limit result $N_0(E_F)$ for the long-range disorder in Fig.~\ref{fig3}(c), while it is a good approximation from the short range disorder in Fig.~\ref{fig3}(d).
The large level broadening for the long-range disorder also implies that  the HLLs-assisted backscattering  becomes important here.
Such a larger level broadening is in consistence with the previous studies of the strong-field density of states in disordered two-dimensional electron systems.\cite{Xie1988,Xie1990}
As a result, we conclude that we find a positive magneto-conductivity for the long-range disorder case which can not be described by the
semiclassical Boltzmann transport theory.
In the rest of the paper, we will clarify the underlying physical mechanisms of the positive magneto-conductivity.
\subsection{Mechanisms for Positive Magneto-conductivity}
As we emphasized in Sec.III,  the isotropic long-range disorder ( $\xi_{z}=\xi_{\|}$ ) can reduce to
two special cases of anisotropic long-range disorder  with  $\xi_{z}\rightarrow0$ and $\xi_{\|}\rightarrow0$, respectively.
Therefore, to gain more insight into the isotropic long-range disorder effect,
it is instructive to study these special cases in (ii-b) and (ii-c).

\begin{figure}
\centering
\includegraphics[scale=0.42, bb = 0 0 1100 550, clip=true]{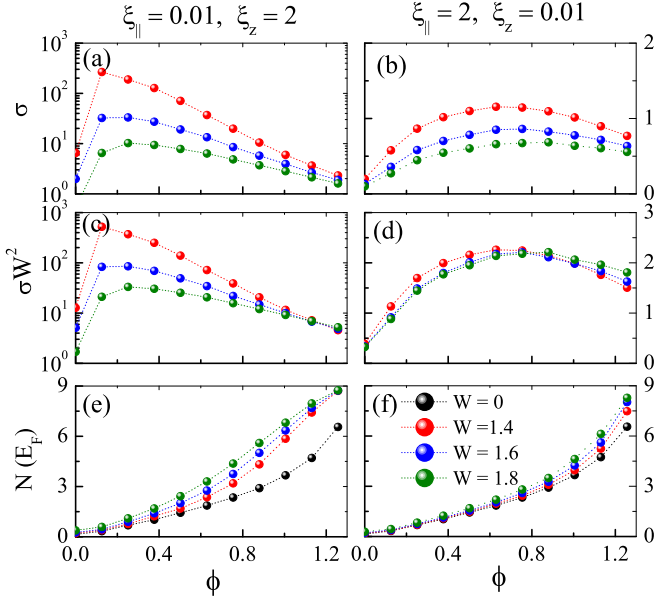}
\caption{(Color online). The conductivity $\sigma$ of the sample at different disorder strength with the long-range disorder (a) $\xi_{\|}=0.01$, $\xi_{z}=2$ and $n_i=0.28$, and (b) $\xi_{\|}=2$, $\xi_{z}=0.01$ and $n_i=0.08$, respectively.
(c) and (d) plot $\sigma W^2$ against the magnetic phase $\phi$ with same data as (a) and (b), respectively.
(e)-(f) the density of states $N(E_F)$ versus $\phi$ at different $W$ corresponding to two types of anisotropic
disorder of (a) and (b), respectively. The size is $20\times50\times100$ and the Fermi energy $E_F=-0.2$.
\label{fig4} }
\end{figure}
\begin{figure}
\centering
\includegraphics[scale=0.42, bb = 0 0 1100 550, clip=true]{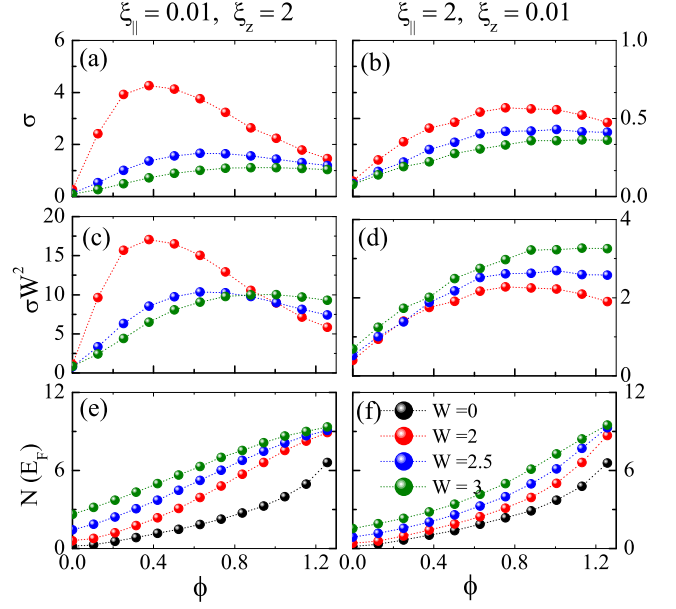}
\caption{(Color online). The conductivity $\sigma$ versus the magnetic phase $\phi$ at different disorder strength $W$ with the long-range disorder (a) $\xi_{\|}=0.01$, $\xi_{z}=2$ and $n_i=0.28$, and (b) $\xi_{\|}=2$, $\xi_{z}=0.01$ and $n_i=0.08$, respectively.
(c) and (d) plot $\sigma W^2$ against $\phi$ with same data as (a) and (b), respectively.
(e)-(f) the density of states $N(E_F)$ versus the magnetic phase $\phi$ at different disorder strength for two type of anisotropic
disorder corresponding to  (a) and (b), respectively.
All the parameters are the same as those of Fig.~\ref{fig4} except the disorder strength $W$.
\label{fig5} }
\end{figure}
Next, let us investigate the weak anisotropic long-range disorder in Fig.~\ref{fig4}.
In Fig.~\ref{fig4}(b), we find a peak of the conductivity for the weak long-range disorder with $\xi_{\|}=2$, $\xi_{z}=0.01$ and $n_i=0.08$. From the Eq.~(\ref{equ:13}), the peak of the conductivity $\sigma$
comes from the interplay between the increasing density of states $N_0(E_F)\propto1/l_B^2$ and the decreasing Fermi velocity $v_F(B)$.
Moreover, the $\sigma W^2$ curves nearly merge [see Fig.~\ref{fig4}(d)], consistent with the semiclassical transport theory.
%because $\sigma W^2\propto N_0{E_F} v_F^2$.
On the other hand, in Fig.~\ref{fig4}(a), we find a large conductivity for the weak  long-range disorder (with $\xi_{\|}=0.01$, $\xi_{z}=2$ and $n_i=0.28$). That is because the long-range correlated disorder in $z$-direction suppresses the large $k_z$ backscattering and thus gives rise to a extremely long transport lifetime in $z$-direction.
In addition, we find that the $\sigma W^2$ curves do not merge in Fig.~\ref{fig4}(c), meaning that the semiclassical theory is not applicable here. The conductivity $\sigma$ decreases (with $W$) more dramatically than the $W^{-2}$ order [see the larger $W$ curves far below the smaller ones in Fig.~\ref{fig4}(c) ].
In fact, in Eq.~(\ref{equ:15}), because the damping factor $\exp(-2k_F^2\xi_z^2)$  from $W_{k_x,k_F;k'_x,-k_F}=\langle|\langle k_F, k_x,  0|V(r)|0, k_x', -k_F \rangle|^2\rangle$ strongly suppresses the direct backscattering $ |0,k_x,k_F\rangle\rightarrow|0,k_x',-k_F \rangle$,
 the HLLs-assisted backscattering ($|0,k_x,k_F \rangle \rightarrow |\nu\neq0, k_x', k_z'\rangle \rightarrow |0, k_x', -k_F\rangle$) becomes dominating due to  the absence of exponential damping term $\exp(-2k_F^2\xi_z^2)$.
As a consequence, we conclude that, for the long-range disorder in $z$-direction, the HLLs-assisted backscattering  is now dominating even at the weak disorder.

If we further increase the disorder strength, the disorder density of states $N(E_F)$ strongly deviated from the clean limit value, meaning that
the energy level broadening at the Fermi energy becomes more apparent [see Fig.~\ref{fig5}(e) and (f)].
Now, the magnetic conductivity $\sigma$ peak (for both types of anisotropic disorder)  moves to a larger magnetic phase $\phi$ with increasing $W$ and $\sigma W^2$ curves do not merge together [see Fig.5(a)-(d)].
Therefore, we conclude that when the level broadening at Fermi level is important for the relative strong disorder strength, the HLLs-assisted scattering plays a dominant role.
In this circumstance, when the Landau level spacings increase  with increasing magnetic phase $\phi$,
the transport lifetime $\tau$ increases because the HLLs-assisted scattering decreases.
Then the conductivity $\sigma=e^2N_0(E_F)v_F^2\tau$ increases with the magnetic phase $\phi$, because $N_0(E_F)$ also increases with $\phi$.
Furthermore, because the HLLs-assisted backscattering (or the transport lifetime $\tau$) is direct related the level broadening (or the disorder strength $W$), it is natural to expect that the conductivity peak would move to a higher magnetic field with increasing disorder strength.
These results and their physical mechanisms are similar to those of Fig.\ref{fig2}(a) for the isotropic long-range disorder .

\section{Discussion and Conclusions}
We have studied transport properties of the WSMs under a strong magnetic field for the short-range and long-range disorder cases, respectively.
The weak short-range disorder can give rise to a monotonous negative magnetic conductivity,
because the Fermi velocity decreases with the magnetic field.
Remarkably, the magnetic conductivity for the long-range disorder can be positive.
For the weak long-range correlated disorder in $z$-direction with, due to the suppressing of large $k_z$ backscattering,
the system has a very long transport lifetime.
In this circumstance, the HLLs-assisted backscattering  becomes dominant and the Boltzmann transport theory is
no longer applicable.
The weak long-range correlated disorder in $x-y$ plane can lead to a positive magnetic conductivity directly, which agrees with the Boltzmann transport theory.
For the relative strong long-range correlated disorder, the HLLs-assisted backscattering  is dominating  and the system has a
conductivity peak moving towards a larger magnetic field with the increasing disorder strength $W$.
Our results have important implications for the recent magneto-transport experiments. In the ultra-quantum limit, the smooth
disorder potential can give rise to a positive magnetic conductivity experimentally observed in the WSMs.
However, such results cannot be obtained from the semiclassical Boltzmann transport theory, because the HLLs-assisted backscattering is always important.
In conclusion, there are two physical mechanisms for the positive magnetic conductivity.
Firstly, the  long-range correlated disorder in $x$-$y$ plane suppresses the intra-Landau-level scattering in the ultra-quantum limit.
Secondly, the probability of the HLLs-assisted backscattering is reduced by increasing the magnetic field.

\section {Acknowledgements.}
This work was financially supported by NBRP
of China (Nos.2015CB921102, 2014CB920901,  and 2012CB921303) and
NSF-China under Grants Nos.11504008, 11534001 and 11374219.

\end{document}